\documentclass[showpacs,twocolumn,prb]{revtex4-1}
\usepackage{graphicx}
\graphicspath{{figures/}} % Location of the graphics files

\newcommand{\be}{\begin{equation}}
\newcommand{\ee}{\end{equation}}
\newcommand{\bea}{\begin{eqnarray}}
\newcommand{\eea}{\end{eqnarray}}
\newcommand{\bwt}{\begin{widetext}}
\newcommand{\ewt}{\end{widetext}}

\usepackage{hyperref}
\hypersetup{
    colorlinks=true,
    linkcolor=blue,
    filecolor=magenta,      
    urlcolor=blue,
}

\usepackage{color} 
\newcommand{\itt}{\it}

\newcommand{\red}{\textcolor{black}}

\usepackage{comment}
\def\comment#1{}

\begin{document}

\title{The possible role of van Hove singularities in the high $T_c$ of superconducting H$_3$S}

\author{Thiago X. R. Souza}
\affiliation{Department of Physics, University of Alberta, Edmonton, AB, Canada T6G~2E1,\\ {\rm and} \\
DComp, Universidade Federal de Sergipe, 49100-000 Sao Cristovao, SE, Brazil}
%Departamento de Fisica, Universidade Federal de Sergipe, 49100-000 Sao Cristovao, SE, Brazil}
\author{F. Marsiglio}
\affiliation{Department of Physics, University of Alberta, Edmonton, AB, Canada T6G~2E1}

\begin{abstract}
We observe that H$_3$S has a bcc structure and, with nearest neighbour hopping only, a strong singularity 
occurs at zero energy. This singularity is accompanied with a highly nested Fermi surface, which is {\it not} conducive
to a stable superconducting instability. Introduction of next-nearest-neighbour hopping removes the singularity, but a `robust'
peak remains in the electron density of states. Solution of the BCS equations shows an enhanced superconducting $T_c$ due to
this peak. Furthermore, nesting is no longer present, so other instabilities will not compete effectively with superconductivity. We 
find high critical temperatures are possible, even with very modest coupling strengths. We also examine a limit of the 
$T=0$ equations (in an Appendix) where an analytical solution is possible over the entire range of coupling strengths, and 
therefore the BCS-BEC crossover is fully covered.
\end{abstract}

\pacs{}
\date{\today }
\maketitle

\section{introduction}

It has been essentially two years since a superconducting transition in the vicinity of $200$ K was first reported in 
hydrogen sulfide.\cite{drozdov14} Since this time, however, experimental results concerning this system have been few,
and to our knowledge, only one as-yet unpublished report has independently confirmed high temperature superconductivity via the
Meissner effect.\cite{huang16} Nonetheless, the crystal structure has now been determined
\cite{einaga16} to be one of two variations of body-centred-cubic (BCC), and is associated with the stoichiometry H$_3$S.
An optical spectroscopy study has also appreared,\cite{capitani16} which claims to provide significant support for an
electron-phonon-based mechanism for superconductivity.

Much of the work to date on this compound has been on the theoretical side. Remarkably, 
even before the experimental discovery of superconducting
H$_3$S, a Density Functional Theory (DFT) calculation \cite{duan14} predicted the correct high pressure structure, and a crude
estimate based on the Allen-Dynes-McMillan formula\cite{mcmillan68} suggested $T_c \approx 200$ K.
Follow-up DFT calculations confirmed this work.\cite{duan15,errea15,papa15,bianconi15,flores-livas15} Several of
these authors furthermore emphasized the
electron-phonon interaction as the mechanism for superconductivity, primarily through the high frequency optical modes
affiliated with the hydrogen atoms. These authors disagree, however, on the importance of anharmonicity, with Errea et al. and
Papaconstantopoulos et al. finding evidence for large anharmonic effects, while Flores-Livas et al. do not. 

In the meantime, Hirsch and one of the present authors \cite{hirsch15} have suggested that it is the conduction by holes through
the sulfur ions that plays a primarily role in the superconductivity. The theoretical framework for the mechanism involved is
expanded upon in earlier work,\cite{hirsch89,marsiglio90} and will not be further discussed here.

The point we wish to make in this paper is that, somewhat independent of the mechanism, a large density of states near the
Fermi level will enhance superconducting $T_c$.  This point has been made repeatedly in the past, starting with the A15
compounds in the 1960's and continuing with the cuprates over the past three decades. A survey of the effects of van Hove
singularities in two and three dimensions on superconducting $T_c$ was published recently.\cite{souza16} Here we wish
to emphasize that the three dimensional BCC structure, pertinent to superconducting H$_3$S, has a logarithmic (squared)
singularity in the density of states when only nearest-neighbour hopping is taken into account, and this has a
significant impact on superconducting properties.\cite{souza16} This was already recognized long ago by Jelitto.\cite{jelitto69}
As already discussed in Ref. [\onlinecite{souza16}], a singularity also exists for the (face-centred-cubic) FCC structure, and in
fact occurs at a filling where nesting effects [which favour other instabilities (e.g. charge density waves)] are not present. 
We will focus on the BCC structure in this paper, and maintain
a non-zero next-nearest neighbour hopping probability, as this seems to more accurately describe the actual situation in H$_3$S; it also
serves to eliminate deleterious effects due to nesting, that would occur in the nearest-neighbour hopping only case for the BCC
structure.

\section{The BCS formalism}

The BCS equations are \cite{schrieffer64,tinkham96}
\be
\Delta_k = -{1 \over N} \sum_{k^\prime} V_{kk^\prime} { \Delta_{k^{\prime}} \over 2 E_{k^\prime}} \left[ 1 - 2f(E_{k^\prime}) \right],
\label{bcs1}
\ee
and
\be
n = {1 \over N} \sum_{k^\prime} \left[ 1 - {\epsilon_{k^\prime} - \tilde{\mu} \over E_{k^\prime}} \left( 1 - 2f(E_{k^\prime})\right) \right],
\label{bcs2}
\ee
with
\be
E_k \equiv \sqrt{  (\epsilon_{k^\prime} - \tilde{\mu})^2 + \Delta^2_{k^\prime}}.
\label{bcs3}
\ee
Here, the wave vector summations cover the First Brillouin zone (FBZ), and we focus on a single band, whose characteristics
are contained within $\epsilon_k$. Similarly the pairing potential, $V_{k,k^\prime}$ is specified by the model under consideration,
and the chemical potential, $\mu$, gives us the density of electrons, $n$. In practice, we `know' the electron density, $n$, and 
therefore need to determine the chemical potential that leads to the desired electron density, for a particularly pairing potential
and temperature (as included through the Fermi-Dirac distribution function,
$f(x) \equiv 1/[{\rm exp}(\beta x) + 1]$, where $\beta \equiv 1/[k_BT]$ is the inverse temperature, with $k_B$ the 
Boltzmann constant). In Eqs. (\ref{bcs1}-\ref{bcs3}), we use $\tilde{\mu}$, which is assumed to include corrections to $\mu$ 
associated with the normal state.

In what follows we assume a featureless attractive interaction, denoted as $V_{k,k^\prime} = -V$, with $V > 0$. This model
constitutes the so-called attractive Hubbard model, as a constant in wave vector space implies an onsite attraction only.
As discussed by Eagles,\cite{eagles69}, Leggett,\cite{leggett80} and Nozi\`eres and Schmitt-Rink,\cite{nozieres85} these 
equations are valid for all pairing strengths (at $T=0$); we discuss a particular limit in the Appendix where these equations
can be solved exactly. Here in the main text, we introduce a cutoff for the pairing potential, so that attraction occurs only for states within
an energy $\hbar \omega_D$ of the Fermi energy, i.e.
\be
V_{kk^\prime} = -V \theta \left[ \hbar \omega_D - |\epsilon_k - \mu | \right] \theta \left[ \hbar \omega_D - |\epsilon_{k^\prime} - \mu | \right] 
\label{vpair}
\ee
where $\theta[x]$ is the  Heaviside step function. Note that removal of this restriction reduces this model to the usual attractive
Hubbard model; identification of $\omega_D$ with a phonon energy scale follows the original BCS treatment, though a more
accurate procedure would be to use the Eliashberg equations,\cite{eliashberg60,marsiglio08} where retardation effects
are properly accounted for. We note that Sano et al.\cite{sano16} have already done for H$_3$S.

The main purpose of this paper is to highlight the importance of electronic structure, through peaks in the electronic
density of states (EDOS) for superconducting $T_c$. Both Quan and Pickett,\cite{quan16} and Sano et al.\cite{sano16}
have included and highlighted this point, based on the results of DFT calculations. 
In our previous work\cite{souza16} we have focused on simple tight-binding descriptions, where, in our opinion, 
the origin of the peak in the density of states is more transparent.

We utilize the BCC; including both nearest and next-nearest neighbour hopping parameters results in the
dispersion
\bea
\epsilon_k &=& -8t\left[ {\rm cos}({k_xa \over 2})  {\rm cos}({k_ya \over 2}) {\rm cos}({k_za \over 2}) \right] \phantom{aa} {\rm [bcc \ \ NNN]} \nonumber \\
& & -2t_{2}\left[ {\rm cos}(k_xa) + {\rm cos}(k_ya) +{\rm cos}(k_za) \right],
\label{bcc_dispersion}
\eea
where $t$ and $t_2$ are the nearest and next-nearest neighbour hopping amplitudes, respectively. The only real impact on the BCS
equations is most readily seen by rewriting them as follows (we also replace the pairing potential $V_{k,k^\prime} = -V$ and linearize
the equations so that they are valid only at $T=T_c$),
\be
{1 \over V} = \int_{\mu_-}^{\mu_+} \ d\epsilon g(\epsilon) {{\rm tanh}[\beta_c (\epsilon -\mu)/2] \over 2(\epsilon - \mu)} \phantom{aaaaa} [T=T_c]
\label{bcs1tc}
\ee
and
\be
n = 2\int_{\epsilon_{\rm min}}^{\epsilon_{\rm max}}   \ d\epsilon g(\epsilon) f(\epsilon - \mu),  \phantom{aaaaaaaaaa}  [T=T_c]
\label{bcs2tc}
\ee
where only the EDOS, $g(\epsilon)$, contains information about the structure. Here $\beta_c \equiv 1/[k_BT_c]$. 
The integration limits in Eq.~(\ref{bcs1tc}) are normally $\mu_- = \mu - \hbar \omega_D$
and $\mu_+ = \mu + \hbar \omega_D$, but when $\mu$ is close to a band edge, then these limits are given by
$\mu_- \equiv{\rm max}[ \mu - \hbar \omega_D, \epsilon_{\rm min}]$, and 
$\mu_+ \equiv{\rm min}[ \mu + \hbar \omega_D, \epsilon_{\rm max}]$, where $\epsilon_{\rm min}$ ($\epsilon_{\rm max}$)
is the energy of the bottom (top) of the band.

\bigskip

\section{Results}

\subsection{The BCC electronic density of states}

\begin{figure}[tp]
\begin{center}
\includegraphics[height=3.4in,width=2.8in,angle=-90]{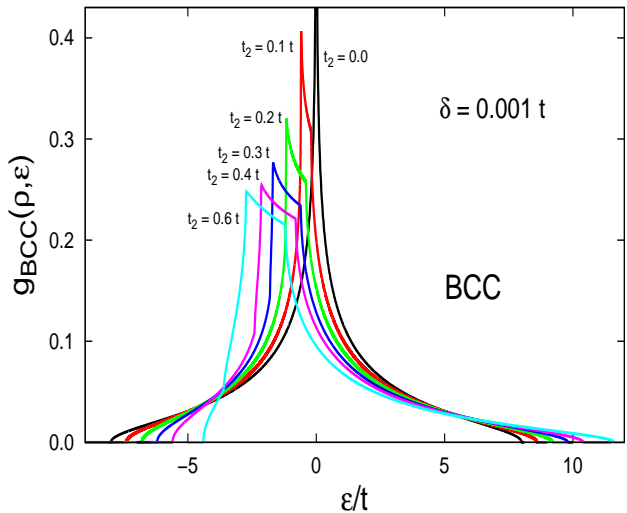}
\end{center}
\caption{Plot of the tight-binding 3D BCC EDOS for different values of the $nnn$ hopping parameter, $t_2$, with $\rho \equiv t_2/t$.
Note that the singularity for $\epsilon = 0$ disappears as $t_2$ becomes non-zero. Nonetheless a large peak, displaced from
$\epsilon = 0$, remains in its place. Results are shown for negative $t_2$
since the results from DFT indicate a structure in the EDOS very similar to this one.\cite{quan16} Moreover, for positive values 
of $\rho$ the results are symmetric (about $\epsilon = 0$) to those 
shown. We used $\delta = 0.001 t$ to generate these results using Eq.~(\ref{num_dense}) [the result for $t_2 = 0$ is indistinguishable
from the more accurate result given by Eq.~(\ref{bcc_anal})].}
\label{fig1bled}
\end{figure}

The EDOS for the BCC structure with nearest-neighbour (nn) hopping only is given by\cite{jelitto69,souza16} 
\be
g_{\rm BCC}(\epsilon) = {2 \over a^3} {1 \over 2 \pi^3 t}\int_{|\bar{\epsilon}|}^1 dx \ {1 \over \sqrt{x^2 - \bar{\epsilon}^2}} K\left[1-x^2\right].
\label{bcc_anal}
\ee
where $\bar{\epsilon} \equiv \epsilon/(4t)$, and $K(z)$ is the complete elliptic integral of the first kind.\cite{olver10}
This function diverges logarithmically at $z\rightarrow 0$, and results in
\be
\lim_{\bar{\epsilon} \rightarrow 0} g_{\rm BCC}(\epsilon) \approx {\ln}^2({1 \over |\bar{\epsilon}|}),
\label{asym}
\ee
which is a stronger divergence than occurs in two dimensions. When next-nearest-neighbour (nnn) hopping is included, then we use
a limiting representation for the $\delta$-function and determine the EDOS through
\be
g_\delta(\epsilon) = {1 \over 2 t a^3} {1 \over \sqrt{\pi \delta^2}} \int_0^1  dx \int_0^1 dy \int_0^1 
dz \ e^{-\left[{\epsilon - \epsilon_k \over 2t\delta}\right]^2},
\label{num_dense}
\ee
where we have substituted $x \equiv k_x a/\pi$ and similarly for $y$ and $z$, and $\delta$ is some small numerical 
smearing parameter (e.g. $\delta = 0.0005t$). In fig.~\ref{fig1bled} we show the BCC density of states for a variety 
of values of $t_2/t$. Note how the singularity evolves (and disappears) once $t_2/t$ departs from zero. Nonetheless, a highly
peaked structure remains for modest values of $t_2/t$.

% fig. 1bled

\subsection{$T_c$}

\begin{figure}[tp]
\begin{center}
\includegraphics[height=4.0in,width=3.6in,angle=-90]{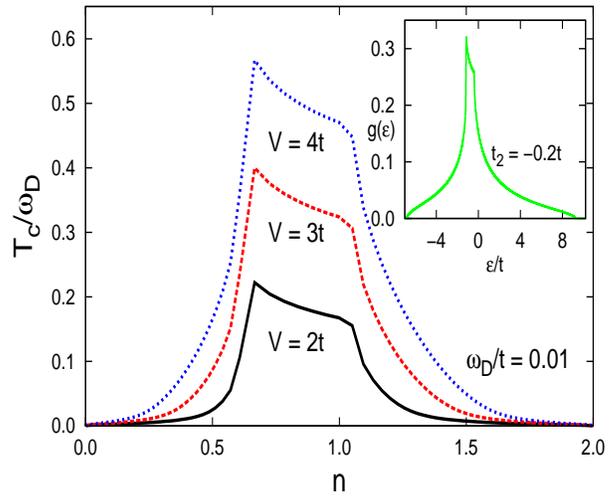}
\end{center}
\caption{\red{Plot of $T_c/\omega_D$ vs. electron density $n$ for  various values of coupling strength, $V/t = 2$, $3$, and $4$. We have use a value of $t_2 = -0.2 t$; the EDOS is plotted in the insert, and resembles very closely the result obtained using DFT.\cite{quan16}
As an example, with  $V = 2t$ ($V/(16t) = 0.125$), and $\omega_D = 100$ meV, 
then $T_c \approx 200$ K (at $n\approx 1$). In this range of $\omega_D$ the results for $T_c$ scale with $\omega_D$.}}
\label{fig2bled}
\end{figure}

To determine $T_c$ one must insert the EDOS from Eq.~(\ref{bcc_anal}) or Eq.~(\ref{num_dense}) into Eqs.~(\ref{bcs1tc},\ref{bcs2tc}),
and perform the ensuing integrals numerically. Based on weak coupling, it is natural to examine dimensionless
quantities, such as $T_c/(\hbar \omega_D)$, vs. $V/t$, $\hbar \omega_D/t$, and $n$. In Fig.~\ref{fig2bled} we show 
$T_c/(\hbar \omega_D)$ as a function of electron density, $n$, for various values of $V$ as indicated. We use $\omega_D = 0.01t$
for definiteness, although this ratio will vary with the specific mechanism that one has in mind. For these values of coupling
strength the ratio $T_c/(\hbar \omega_D)$ is fairly insensitive to $\omega_D/t$, and so this figure can be used for other
values of $\omega_D$ to estimate $T_c$ in real units. This is indicative of weak coupling, so in fact the shape of $T_c$ vs $\mu$
will resemble closely the density of states (as a comparison with the relevant curve in Fig.~\ref{fig1bled} indicates). Here there will
be some distortion since $T_c$ is plotted vs. $n$ and not versus chemical potential.

% fig. 2bled

\section{Summary}

This study does not directly address the mechanism for superconductivity in H$_3$S. Instead, we have found, as have other DFT-based
studies, that the BCC structure itself will tend to amplify pairing effects, due to the possibly very high electronic density of states
at the Fermi level. More generally, the superconducting community should be more aware that singularities in the electronic
densities of states can occur in three dimensions as well as lower dimensions, in all three types of cubic structures, simple, 
face-centred, and body-centred cubic. The existence of this possibility was first pointed out by Jelitto,\cite{jelitto69} and we
elaborated on the nearest-neighbour models considered by him to those that include $nnn$ hopping as well.\cite{souza16}
Although not addressed here, it is also worth noting that the isotope effect is expected to display some peculiar characteristics,
again due to the presence of van Hove singularities in the EDOS.\cite{souza16}

When $nnn$ hopping is introduced, the singularity disappears in the EDOS.\cite{remark1} The peak that remains is in some ways more
`robust' --- it (and therefore superconducting $T_c$) will withstand more readily the degradation 
that is inevitable due to impurities and imperfections. Note that the realization that the presence of a BCC structure in the 
material will lead to an enhanced $T_c$ occurred also through DFT studies. Nonetheless, it is beneficial to have simplified 
tight-binding models like the one presented here to help identify important structure characteristics for enhancing $T_c$.
It is clear from characteristics of the EDOS, that doping with electrons (should that become possible) will lead to a lower $T_c$ if
this is all that mattered. Some mechanisms (e.g. the "hole" mechanism\cite{hirsch15}) predict a strong doping dependence
independently of changes in the EDOS, and then the qualitative prediction of this model will depend on whether H$_3$S lies on the
electron- or hole-side of the maximum predicted in that model. The dependence of the effective interaction on doping is expected
to overwhelm the dependence of the EDOS on doping in this particular model. It will be interesting to see if such experiments can
be carried out.

\begin{acknowledgments}

This work was supported in part by the Natural Sciences and Engineering
Research Council of Canada (NSERC). TXRS is a recipient of an "Emerging Leaders in the Americas Program" (ELAP) scholarship 
from the Canadian government, and we are grateful for this support.

\end{acknowledgments}

\appendix

\section{Exact results (within $T=0$ BCS) for any filling and coupling strength for a constant density of states}

%see Nov. 29, 2016 written notes for a full derivation
We start with Eqs.~(\ref{bcs1}-\ref{bcs2}), and illustrate that for the attractive Hubbard model,
\be
V_{k,k^\prime} = -|U|,
\label{att_hubb_pot}
\ee
an exact solution exists\cite{remark2} at $T=0$, if we adopt a constant density of states model, $g(\epsilon) = 1/W$, where
$W$ is the electronic bandwidth, and the band extends from $-W/2$ to $W/2$. 
With these assumptions, Eqs.~(\ref{bcs1}-\ref{bcs2}) become
\bea
{2 W \over |U|} &=& \int_{-W/2}^{+W/2} \ d\epsilon { 1 \over \sqrt{(\epsilon - \tilde{\mu})^2 + \Delta^2}} \label{bcs_flat1} \\
W(1-n) &=& \int_{-W/2}^{+W/2} \ d\epsilon { \epsilon - \tilde{\mu} \over \sqrt{(\epsilon - \tilde{\mu})^2 + \Delta^2}},
\label{bcs_flat2} 
\eea
where $\tilde{\mu} \equiv \mu + |U|n/2$ is the bare chemical potential with the Hartree correction.
As remarked in Ref. [\onlinecite{nozieres85}], these equations amount to a change of variables from ($\Delta,\tilde{\mu}$)
to ($|U|,n$). These equations retain their validity for all coupling strengths, from weak to strong coupling, and describe 
pairing from the Cooper pair limit to the Bose Condensed pair limit. To our knowledge, they have never been inverted 
analytically over the entire range of parameters until now.

To proceed, one performs both (elementary) integrals. Successive squaring of the result from Eq.~(\ref{bcs_flat2}) results
in an explicit determination of $\tilde{\mu}$ in terms of $\Delta$:
\be
\tilde{\mu} = -{W \over 2}(1-n) \left\{1 + \left({\Delta \over W/2}\right)^2{1 \over n (2-n)}\right\}. 
\label{mu_eqn}
\ee
Proceeding with Eq.~(\ref{bcs_flat1}), defining variables $Y \equiv {\rm exp}(2W/|U|)$ and $x \equiv \tilde{\mu} + W/2$,
successive squaring of this equation results in yet another explicit determination of $\tilde{\mu}$:
\be
\tilde{\mu}^2 = \left({W \over 2}\right)^2 \left( {Y + 1 \over Y - 1} \right)^2 \left\{1 - \left({\Delta \over W}\right)^2{(Y-1)^2 \over Y}\right\}. 
\label{mu_eqn2}
\ee
Equating Eq.~(\ref{mu_eqn2}) with the square of Eq.~(\ref{mu_eqn}) then allows us to solve for $\Delta$; back-substituting
this result into Eq.~(\ref{mu_eqn}) then gives us an explicit result for $\mu$. The final results are
\be
\mu = -{|U| \over 2} n - {W \over 2} (1-n) {\rm coth}\left({W \over |U|}\right),
\label{mu_explicit}
\ee
and
\be
\Delta = {W \over 2} \sqrt{ { n(2-n) \over 1 - (1-n){\rm tanh}^2\left({W \over |U|}\right)} } {\rm csch}\left({W \over |U|}\right).
\label{gap_explicit}
\ee
From Eq.~(\ref{mu_explicit}) one immediately obtains
\be
n = {\mu + {W \over 2} {\rm coth}\left({W \over |U|}\right) \over {W \over 2} {\rm coth}\left({W \over |U|}\right) - {|U| \over 2}};
\label{occup}
\ee
this result includes both Hartree and pairing contributions.
One can readily verify that the weak and strong coupling limits are achieved correctly with these equations.


\begin{thebibliography}{99}
\bibitem{remark1} This is true for the BCC and FCC lattice structures but, as discussed in Ref. [\onlinecite{souza16}],
the introduction of $nnn$ hopping in the SC model will actually create a singularity in the EDOS.
\bibitem{remark2} Again we emphasize that this statement is true within BCS theory for a finite band with a constant
density of states. To our knowledge even with these caveats, this solution has not been noticed before in the literature. 
A truly exact solution is available by
way of the Bethe Ansatz only in one dimension,\cite{lieb68} and comparisons have been made with BCS theory in
Ref. [\onlinecite{marsiglio97}].
\end{thebibliography}
\end{document}